\documentclass[final,5p,times,twocolumn]{elsarticle}

\usepackage[T1]{fontenc}
\usepackage[utf8]{inputenc}

\usepackage{amsmath}
\usepackage{amsfonts}
\usepackage{amssymb}
\usepackage{amsxtra}
\usepackage{array}
\usepackage{graphicx}
\usepackage{hepunits}
\usepackage{units}
\usepackage{color}
\usepackage{xspace}

\definecolor{purple}{rgb}{0.5,0,0.5}
\definecolor{blue}{rgb}{0.0,0,0.9}
\definecolor{prdblue}{rgb}{0.133,0.118,0.498}
\usepackage[colorlinks=true, pdfstartview=FitV, linkcolor=prdblue, citecolor= prdblue, urlcolor=prdblue]{hyperref}

\usepackage[mathscr,scaled=1.15]{urwchancal}
\DeclareFontFamily{OT1}{pzc}{}
\DeclareFontShape{OT1}{pzc}{m}{it}%
{<-> s * [1.15] pzcmi7t}{}
\DeclareMathAlphabet{\mathpzc}{OT1}{pzc}{m}{it}

\biboptions{sort&compress}

\journal{Physics Letters B}

\begin{document}

\begin{frontmatter}

\title{$\,$\\[-7ex]\hspace*{\fill}{\normalsize{\sf\emph{Preprint no}. NJU-INP 055/22}}\\[1ex]
Parton distributions of light quarks and antiquarks in the proton}

\author[NKU]{Lei Chang} 

\author[PKU]{Fei Gao} 

\author[NJU,INP]{Craig D. Roberts}

%
\address[NKU]{
School of Physics, Nankai University, Tianjin 300071, China}
\address[PKU]{
Centre for High Energy Physics, Peking University, Beijing 100871, China}
\address[NJU]{
School of Physics, Nanjing University, Nanjing, Jiangsu 210093, China}
\address[INP]{
Institute for Nonperturbative Physics, Nanjing University, Nanjing, Jiangsu 210093, China\\[1ex]
Email addresses:
\href{mailto:leichang@nankai.edu.cn}{leichang@nankai.edu.cn} (L. Chang);
\href{mailto:hiei@pku.edu.cn}{hiei@pku.edu.cn} (F. Gao);
\href{mailto:cdroberts@nju.edu.cn}{cdroberts@nju.edu.cn} (C. D. Roberts);
}

\begin{abstract}
An algebraic \emph{Ansatz} for the proton's Poincar\'e-covariant wave function, which includes both scalar and pseudovector diquark correlations, is used to calculate proton valence, sea, and glue distribution functions (DFs).  Regarding contemporary data, a material pseudovector diquark component in the proton is necessary for an explanation of the neutron-proton structure function ratio; and a modest Pauli blocking effect in the gluon splitting function is sufficient to explain the proton's light-quark antimatter asymmetry.  In comparison with pion DFs, the light-front momentum fractions carried by all identifiable parton classes are the same; on the other hand, the higher moments are different.  Understanding these features may provide insights that explain distinctions between Nambu-Goldstone bosons and seemingly less complex hadrons.
\end{abstract}


\end{frontmatter}

\noindent\textbf{1.$\;$Introduction}.
The proton is Nature's most fundamental bound-state and efforts to elucidate its structure have a one-hundred-year history.  A variety of probes have been used to good effect, e.g., electron scattering provides access to both elastic form factors \cite{Arrington:2006zm, Punjabi:2015bba, Barabanov:2020jvn, Holt:2012gg} and inelastic structure functions \cite{Holt:2012gg, Holt:2010vj, Hen:2016kwk}, and also an array of Drell-Yan reactions \cite{Holt:2012gg, Holt:2010vj, Reimer:2007iy, Peng:2014hta, Geesaman:2018ixo}.  A primary aim of such efforts is to understand how the gluon and quark fields used to express the Lagrangian of quantum chromodynamics (QCD) interact and combine to form the proton.  Given the empirical fact of confinement \cite{Krein:1990sf, Jaffe:Clay, Papavassiliou:2016anw, Gao:2017uox, Binosi:2019ecz} and issues relating to the emergence of hadron masses \cite{Roberts:2016vyn, Krein:2020yor}, this is a very difficult problem.  Of course, in order to solve QCD, more is required than a realistic sketch of proton structure.  Excited states of the proton must be understood \cite{Burkert:2017djo, Brodsky:2020vco, Carman:2020qmb}; and mesons, a potentially distinct form of hadron matter, especially the Nambu-Goldstone bosons (pions and kaons) \cite{Horn:2016rip, Aguilar:2019teb, Chen:2020ijn, Arrington:2021biu, Roberts:2021nhw}.  Nevertheless, the proton's stability has made it the focus of most attention.

Since the discovery of quarks in deep inelastic scattering \cite{Kendall:1991np, Taylor:1991ew, Friedman:1991nq, Friedman:1991ip} and the associated development of the quark-parton model \cite{Feynman:1969wa, Bjorken:1969ja} and QCD \cite{Marciano:1977su}, much has been learnt about proton structure using data that can be related to parton distribution functions (DFs), \emph{viz}.\ number density distributions of gluons and quarks.
Of special interest are possible differences between the DFs of the proton's $d$- and $u$-quarks, ${\mathpzc d}(x)$, ${\mathpzc u}(x)$, where $x$ is the fraction of the proton's light-front momentum carried by the struck quark.  This is because the value of ${\mathpzc d}(x)/{\mathpzc u}(x)$, on the far-valence domain ($x\simeq 1$) is a keen discriminator between competing descriptions of proton structure \cite{Roberts:2013mja, CLAS:2014jvt, Abrams:2021xum}; and because of the potential impact of any such differences on the proton's antimatter (sea quark) DFs \cite{Field:1976ve, NewMuon:1991hlj, NewMuon:1993oys, NA51:1994xrz, NuSea:2001idv, SeaQuest:2021zxb}.

An explicitly Poincare\'-covariant three-body treatment of such issues would be best \cite{Eichmann:2016yit, Qin:2020rad}, but that is currently impossible.  Herein, therefore, we address both questions using a quark+diquark representation of the proton's Poincar\'e-covariant wave function \cite{Cahill:1988dx, Reinhardt:1989rw, Efimov:1990uz}, in which both $[ud]$ (isoscalar-scalar -- $0^+$) and $\{uu\}$, $\{ud\}$ (isovector-pseudovector -- $1^+$) correlations can be present.  The diquarks are both nonpointlike and interacting.  This approach has met with success in widespread applications to baryon spectroscopy and elastic and transition form factors \cite{Barabanov:2020jvn}.  Hitherto, however, only very simple formulations have been used in analyses of DFs and the applications have largely been restricted to the valence-quark sector \cite{Barabanov:2020jvn}.

\medskip

\noindent\textbf{2.\,Valence quark distributions}.
Suppose a hadron-scale ($\zeta=\zeta_{\cal H}$) proton, with total momentum $K$, is effectively a two-body system, in some practical respects, constituted from mass-degenerate dressed valence-quarks and interacting, nonpointlike isoscalar-scalar and isovector-pseudovector quark+quark (diquark) correlations.  Such a proton can be represented by a canonically-normalised Faddeev amplitude:
\begin{equation}
\label{FAproton}
\psi(\ell;K) = \sum_{J^P=0^+,1_{\{uu\}}^+,1_{\{ud\}}^+} a_{J^P} \psi^{J^P}(\ell;K)\,.
\end{equation}
Here, $a_{0^+},a_{1_{\{uu\}}^+},a_{1_{\{ud\}}^+}$, with $a_{1^+} := a_{1_{\{ud\}}^+} = -a_{1_{\{uu\}}^+}/\surd 2$ owing to isospin Clebsch-Gordon couplings, measure the relative strength of the scalar and pseudovector diquark terms, $\psi^{0^+,1^+}(\ell;K)$ describe the dependence of the amplitude on the quark momentum, $\ell$, and we have suppressed the Lorentz index associated with the $1^+$ correlations.
In this case, one may readily generalise the results in Refs.\,\cite{Chang:2014lva, Chen:2016sno} to obtain expressions for the proton valence-quark DFs.

For instance, ${\mathpzc u}_V^p(x;\zeta_{\cal H})$ receives three distinct contributions:
\begin{equation}
{\mathpzc u}_V^p(x;\zeta_{\cal H}) = \sum_{t=Q,D_0,D_1} {\mathpzc u}_{Vt}^p(x;\zeta_{\cal H})\,.
\label{Equproton}
\end{equation}
The first is provided by the $u$-quark that is not sequestered within a diquark
[$\hat\delta_n^{xK} = n\cdot K \delta(n\cdot \ell - x n\cdot K)$]:
\begin{align}
\Lambda_+ & \gamma\cdot n  {\mathpzc u}_{VQ}^p(x;\zeta_{\cal H}) \Lambda_+   =
\int\! \tfrac{d^4\ell}{(2\pi)^4} \,
\hat\delta_n^{xK} \;
\Lambda_+ Q_u  \Lambda_+\,, \label{uproton1}
\end{align}
with
$\Lambda_+ =(-i\gamma \cdot K+m_p)/(2m_p)$, $K^2=-m_p^2$, $m_p=0.94\,$GeV being the proton mass; $n^2 =0$; and
\begin{align}
Q_u & = \sum_{J^P=0^+,1_{\{uu\}}^+,1_{\{ud\}}^+} \!\! a_{J^P}^2 \,
\bar\psi^{J^P}(\ell-\tfrac{1}{2}K;-K)  \nonumber \\
& \quad \times S(\ell) (-i)\gamma\cdot n S(\ell) \psi^{J^P}(\ell-\tfrac{1}{2}K;K) \Delta^{J^P}(\ell-K)\,,
\label{EqQ}
\end{align}
where $S(\ell)$ is the propagator of the dressed valence-quark and $\Delta^{J^P}(\ell-K)$ are propagators for the diquark correlations.  

The second exposes the $u$-quark within the scalar diquark and is expressed in a convolution:
\begin{align}
{\mathpzc u}_{V D_0}^p(x;\zeta_{\cal H}) = a_{0^+}^2 \int_x^1\,dy\,{\mathpzc s}_{0^+}^p(y;\zeta_{\cal H})
{\mathpzc u}_V^{0^+}(x/y;\zeta_{\cal H})\,, \label{convolution0}
\end{align}
where ${\mathpzc u}_V^{0^+}(x;\zeta_{\cal H})$ is the DF of a valence $u$-quark in a scalar diquark and the probability density for finding a scalar diquark carrying a light-front fraction $x$ of the proton's momentum is
\begin{subequations}
\begin{align}
\Lambda_+ & \gamma\cdot n  {\mathpzc s}_{0^+}^p(x;\zeta_{\cal H}) \Lambda_+   =
\int\! \tfrac{d^4\ell}{(2\pi)^4} \,
\hat\delta_n^{xK} \;
\Lambda_+ D_{0}  \Lambda_+\,,\\
D_{0} & =  n\cdot \partial^\ell
\left[\bar\psi^{0^+}(\ell-\tfrac{1}{2}K;-K)
\underline{S(\ell-K)} \psi^{0^+}(\ell-\tfrac{1}{2}K;K) \Delta^{0^+}(\ell)\right].
\end{align}
\end{subequations}
Here, the underlined propagator is \underline{not} differentiated; so, $D_{0}$ has three distinct terms.

The final piece reveals the $u$-quarks in the pseudovector diquarks, whose form in the isospin-symmetry limit follows:
\begin{subequations}
\label{convolution1}
\begin{align}
{\mathpzc u}_{V D_1}^p(x;\zeta_{\cal H}) &= 5 {\mathpzc q}_{V D_1}^p(x;\zeta_{\cal H})\,, \\
{\mathpzc q}_{V D_1}^p(x;\zeta_{\cal H})& = a_{1^+}^2 \int_x^1\,dy\,{\mathpzc s}_{1^+}^p(y;\zeta_{\cal H})
{\mathpzc u}_V^{1^+}(x/y;\zeta_{\cal H})\,,
\end{align}
\end{subequations}
where the ``5'' owes to isospin Clebsch-Gordon couplings
and the fact that the $\{uu\}$ contains two active valence $u$-quarks, and
\begin{subequations}
\label{uproton3}
\begin{align}
\Lambda_+ & \gamma\cdot n  {\mathpzc s}_{1^+}^p(x;\zeta_{\cal H}) \Lambda_+   = \hat\delta_n^{xK}\;
\Lambda_+ D_{1}  \Lambda_+\,,\\
D_{1} & =  n\cdot \partial^\ell
\left[\bar\psi_\sigma^{1^+}(\ell-\tfrac{1}{2}K;-K)
\underline{S(\ell-K)} \psi_\rho^{1^+}(\ell-\tfrac{1}{2}K;K) \Delta_{\sigma\rho}^{1^+}(\ell)\right]\,.
\end{align}
\end{subequations}
Again, the underlined propagator is \underline{not} differentiated; so, $D_{1}$ also has three distinct terms.

As indicated by Eq.\,\eqref{Equproton}, the final result for ${\mathpzc u}_V^p(x;\zeta_{\cal H})$ is obtained by combining the outputs from Eqs.\,\eqref{uproton1}\,--\,\eqref{uproton3}.

The valence $d$-quark distribution is similarly composed:
\begin{equation}
{\mathpzc d}_V^p(x;\zeta_{\cal H}) = \sum_{t=Q,D_0,D_1} {\mathpzc d}_{V t}^p(x;\zeta_{\cal H})\,,
\end{equation}
where
\begin{align}
\Lambda_+ & \gamma\cdot n  {\mathpzc d}_{V Q}^p(x;\zeta_{\cal H}) \Lambda_+   =
\int\! \tfrac{d^4\ell}{(2\pi)^4} \,
\hat\delta_n^{xK} \;
\Lambda_+ Q_d  \Lambda_+\,, \label{dproton1}\\
Q_d & = 2 a_{1^+}^2
\bar\psi_\sigma^{1^+}(\ell-\tfrac{1}{2}K;-K)  \nonumber \\
& \quad \times S(\ell) (-i)\gamma\cdot n S(\ell) \psi_\rho^{J^P}(\ell-\tfrac{1}{2}K) \Delta_{\sigma\rho}^{1^+}(\ell-K)\,,
\label{EqQd}
\end{align}
${\mathpzc d}_{V D_0}^p(x;\zeta_{\cal H}) = {\mathpzc u}_{V D_0}^p(x;\zeta_{\cal H})$,
${\mathpzc d}_{V D_1}^p(x;\zeta_{\cal H}) = {\mathpzc q}_{V D_1}^p(x;\zeta_{\cal H})$.

\medskip

\noindent\textbf{3.\,Sum rules}.
The electromagnetic current for a proton represented by Eq.\,\eqref{FAproton} may be obtained from that described, \emph{e.g}., in Ref.\,\cite{Segovia:2014aza}.  Using this current and recognising that canonical normalisation of the Faddeev amplitude ensures the proton's Dirac form factors satisfy $F_1^{p,u}(Q^2=0)=2$, $F_1^{p,d}(Q^2=0)=1$, then one may readily establish the following:
\begin{subequations}
\label{baryonnumber}
\begin{align}
\nonumber
 \int_0^1 dx\, & \Lambda_+  \gamma\cdot n {\mathpzc u}_V^p(x;\zeta_{\cal H}) \Lambda_+
 =  \Lambda_+ \gamma\cdot n F_1^{p,u}(0) \Lambda_+ = 2 \Lambda_+ \gamma\cdot n \Lambda_+\\
\Rightarrow &
\int_0^1 dx\, {\mathpzc u}_V^p(x;\zeta_{\cal H})  = 2\,;\\
\mbox{similarly} & \int_0^1 dx\, {\mathpzc d}_V^p(x;\zeta_{\cal H})  = 1\,.
\end{align}
\end{subequations}
%
Given that $F_1^{p,(u,d)}$ are observable, then Eqs.\,\eqref{baryonnumber} remain true $\forall \zeta > \zeta_{\cal H}$. This is expressed in the fact that QCD (DGLAP) evolution \cite{Dokshitzer:1977sg, Gribov:1971zn, Lipatov:1974qm, Altarelli:1977zs} preserves valence-quark (baryon) number.

Since the hadron-scale proton is constituted solely from dressed valence-$u$ and -$d$ quarks, it is intuitively clear that
\begin{equation}
\label{momentumsumrule}
\langle x \rangle_{{\mathpzc u}_p}^{\zeta_{\cal H}}+
\langle x \rangle_{{\mathpzc d}_p}^{\zeta_{\cal H}} :=
\int_0^1 dx\, x [{\mathpzc u}_V^p(x;\zeta_{\cal H}) +{\mathpzc d}_V^p(x;\zeta_{\cal H})] = 1\,.\\
\end{equation}
Similar statements are established for mesons in Refs.\,\cite{Chang:2014lva, Chen:2016sno, Ding:2019lwe, Cui:2020tdf, Chang:2021utv}.  A proof of Eq.\,\eqref{momentumsumrule} may be constructed by using Eqs.\,\eqref{baryonnumber} and generalising the meson arguments via the following identity and its corollaries:
\begin{subequations}
\label{ndotdzero}
\begin{align}
0 & = \int\! \tfrac{d^4\ell}{(2\pi)^4} \, n\cdot\partial^\ell \Lambda_+ {\mathpzc P}(\ell;K) \Lambda_+ \,,\\
{\mathpzc P}(\ell;K) & =
\sum_{J^P=0^+,1_{\{uu\}}^+,1_{\{ud\}}^+} \!\! a_{J^P}^2 \,
\bar\psi^{J^P}(\ell-\tfrac{1}{2}K;-K)  \nonumber \\
& \qquad \times S(\ell)  \psi^{J^P}(\ell-\tfrac{1}{2}K;K) \Delta^{J^P}(\ell-K)\,.
\end{align}
\end{subequations}

\medskip

\noindent\textbf{4.\,\emph{Ans\"atze} connected with the proton amplitude}.
The hadron-scale valence-quark DFs can immediately be calculated once the dressed propagators and Faddeev amplitudes in, \emph{e.g}., Eq.\,\eqref{ndotdzero}, are specified.  After thirty years of study, all relevant quantities are known.  Following Ref.\,\cite{Mezrag:2017znp}, we proceed by using simple algebraic representations for every element, with each form and their relative strengths based on the results of modern analyses \cite{Chen:2017pse, Cui:2020rmu, Lu:2021sgg, Chen:2021guo, Raya:2021pyr}.  Thus freed from cumbersome numerical analysis, this approach brings a clarity that enables quantitative predictions to be made and insights to be drawn.

The dressed-quark and -diquark propagators are:
\begin{subequations}
\begin{align}
& S(\ell)=(-i\gamma\cdot\ell + M)\sigma_{M}(\ell^2)\,,\; \sigma_M(s) = 1/[s+M^2]\,, \\
& \Delta^{0^+}(\ell)=\sigma_{M_0}(\ell^2)\,, \;
\Delta_{\sigma\rho}^{1^+}(\ell)=\delta_{\sigma\rho}\sigma_{M_1}(\ell^2)\,,
\end{align}
\end{subequations}
%
%
Here, the parameters are the masses: $M$ for the dressed light quarks and $M_{0,1}$ for the diquarks.
Transversality of the pseudovector diquark propagator is implemented indirectly via the related correlation amplitude.

Next, the components of the Faddeev amplitude:
$\hat\sigma_M(s) = M^2\sigma_M(s)$,
$\ell_z^2 = (\ell+ z K/2)^2$,
$\gamma_\rho^K = \gamma_\rho - \gamma\cdot K K_\rho/K^2$,
\begin{subequations}
\label{diquarkamps}
\begin{align}
\psi^{0^+}(\ell;K) & =
{\mathbb I} \int_{-1}^1 dz\,\omega(z)
\hat\sigma_\Lambda(\ell_z^2)^2
,\\
%
\psi_\rho^{1^+}(\ell;K) & =
\tfrac{1}{\surd 3}\gamma_5 \gamma_\rho^K \int_{-1}^1 dz\,\omega(z) \hat\sigma_\Lambda(\ell_z^2)^2,
%
\end{align}
\end{subequations}
where we have used perturbation theory integral representations \cite{Chang:2013nia, Ydrefors:2021dwa}, with
$\omega$ a unit-normalised spectral weight whose presence controls the $\ell \cdot K$ dependence of the amplitude.
Ground-states constituted from two mass-degenerate constituents are described by $\bar\omega(z) = (3/4)(1-z^2)$.  However, in a quark+diquark proton, the effective constituents are not degenerate.  We reflect the mass imbalance by using $\omega(z)=(1-2z)/2$.  The same weight is used for both diquark correlations because one typically finds $(M_1-M_0) / (M_1+M_0) \lesssim 0.1$ \cite{Barabanov:2020jvn}.

In writing Eqs.\,\eqref{diquarkamps}, we have retained only the dominant components of the proton Faddeev amplitude, both of which correspond to $L=0$ in the proton rest frame.  There is one more ($L=1$) $0^+$-diquark component and five additional $1^+$ diquark contributions, but all are negligible in comparison with the respective leading term \cite{Chen:2017pse}.  The large-$\ell^2$ dependence of the retained amplitudes is modelled on that found in numerical solutions of the proton Faddeev equation \cite{Chen:2017pse}.

The remaining elements are ${\mathpzc u}_V^{0^+/1^+}(x;\zeta_{\cal H})$ in Eqs.\,\eqref{convolution0}, \eqref{convolution1}, which represent the probability densities for finding a valence-quark in a scalar/pseudovector diquark.  Structural properties of diquark correlations were explored in Ref.\,\cite{Lu:2021sgg}.  The results therein combined with insights developed in Refs.\,\cite{Cui:2020tdf, Roberts:2021nhw} lead to the following \emph{Ansatz}:
\begin{equation}
{\mathpzc u}_V^{0^+,1^+}(x;\zeta_{\cal H})
= {\mathpzc q}_{01}(x;\zeta_{\cal H})
= {\mathpzc n}_{\mathpzc q} x^2 (1 - x)^2 {\rm e}^{ 20 x (1 - x)-1},
\label{EqqqDF}
\end{equation}
where ${\mathpzc n}_{\mathpzc q}$ ensures unit normalisation.  Considering Ref.\,\cite{Lu:2021sgg}, one sees it is reasonable to use the same function for both diquarks.  The key qualitative features of Eq.\,\eqref{EqqqDF} are symmetry under $x\leftrightarrow(1-x)$ and  compression with respect to the scale free DF form, ${\mathpzc q}_{\rm sf}(x)=30 x^2(1-x)^2$.

The model has four parameters: $M$, $M_{0,1}$, ${\mathpzc r}_{10}=a_{1^+}/a_{0^+}$.  We fix the masses by referring to existing Faddeev equation studies \cite{Chen:2017pse, Cui:2020rmu, Lu:2021sgg, Chen:2021guo, Raya:2021pyr}:
\begin{equation}
\label{modelparameters}
\begin{array}{ccc}
M/{\rm GeV} & M_0/{\rm GeV} & M_1/{\rm GeV} \\
0.4 & 0.78 & 0.92
\end{array}\,,
\end{equation}
$\Lambda = [M_0+M_1]/2$.
%
Regarding ${\mathpzc r}_{10}$, we note that in realistic Faddeev equation solutions its value is largely determined by the ratio $M_0/M_1$, with ${\mathpzc r}_{10} \approx 0.45$ being typical \cite{Mezrag:2017znp}.  This value corresponds to the $0^+$ diquark producing roughly 65\% of the proton's normalisation.  Herein, we instead treat ${\mathpzc r}_{10}$ as ``free'', using it to illustrate the impact of the proton's pseudovector diquark content on its DFs.

\medskip

\noindent\textbf{5.\,Valence-quark DFs at large $\mathbf x$}.
The algebraic simplicity of the \emph{Ans\"atze} in Sect.\,4 enables one to expose the large-$x$ behaviour of the hadron-scale valence-quark DFs.  To begin, the pattern of contributions to ${\mathpzc u}_V^p(x;\zeta_{\cal H})$ and ${\mathpzc d}_V^p(x;\zeta_{\cal H})$ which is recorded in Sect.\,2 shows that both will have the same large-$x$ power so long as $a_{1^+}\neq 0$.
Next, working within the context of contour integration, relevant contributions arise from the simple poles associated with the quark propagators, diquark propagators, and Faddeev amplitude functions.
%

The leading behaviours that survive after combining all these contributions are:
{\allowdisplaybreaks
\begin{subequations}
\begin{align}
{\mathpzc u}_V^p(x;\zeta_{\cal H}) & \stackrel{x \simeq 1}{\sim} (1-x)^3\,, \label{partonmodel}\\
\lim_{x\to 1} \frac{{\mathpzc d}_V^p(x;\zeta_{\cal H})}{{\mathpzc u}_V^p(x;\zeta_{\cal H})} & =
\frac{2  \eta {\mathpzc r}_{10}^2}{1 + \eta {\mathpzc r}_{10}^2}\,,
\label{HoltFormula}
\end{align}
\end{subequations}
where $\eta$ measures the ratio of the integrals in Eqs.\,\eqref{uproton1}, \eqref{dproton1}, whose size is determined by $M_0^2/M_1^2$.  Using the values in Eq.\,\eqref{modelparameters}, $\eta=0.58$ and $M_0^2/M_1^2=0.72$.
With Eq.\,\eqref{partonmodel}, our framework reproduces the quark-parton model prediction \cite{Farrar:1975yb, Brodsky:1979gy, Brodsky:1994kg}, which is confirmed in global fits of relevant data \cite{Ball:2016spl, Courtoy:2020fex}.
Further, Eq.\,\eqref{HoltFormula} recovers the result discussed in Refs.\,\cite{Wilson:2011aa, Roberts:2013mja}.  As already noted, ${\mathpzc d}_V^p$ and ${\mathpzc u}_V^p$ exhibit the same power-law behaviour when ${\mathpzc r}_{10}\neq 0$.  Using a more sophisticated current, which explicitly incorporates quark exchange between scalar diquark correlations, ${\mathpzc d}_V^p(x;\zeta_{\cal H})$ has the same power behaviour even when ${\mathpzc r}_{10}= 0$.  However, the strength of this contribution is weak \cite{Xu:2015kta}; namely, changing the value of $\lim_{x\to 1}{\mathpzc d}_V^p(x;\zeta_{\cal H})/{\mathpzc u}_V^p(x;\zeta_{\cal H})$ by less-than 3\%.
}

\begin{figure}[t]
\vspace*{0.5ex}

\leftline{\hspace*{0.5em}{\large{\textsf{A}}}}
\vspace*{-3ex}
\includegraphics[width=0.42\textwidth]{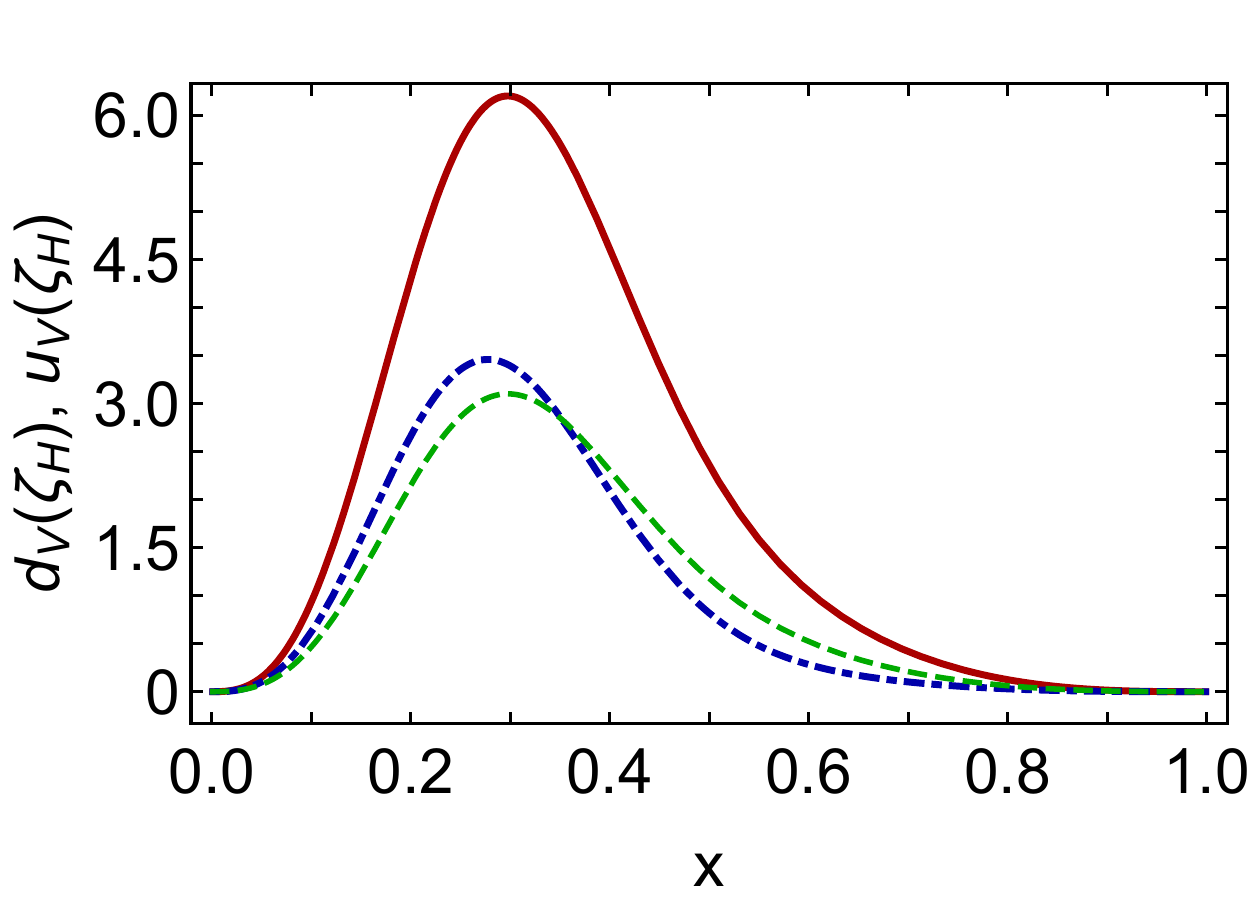}
\vspace*{-1ex}

\leftline{\hspace*{0.5em}{\large{\textsf{B}}}}
\vspace*{-3ex}
\includegraphics[width=0.42\textwidth]{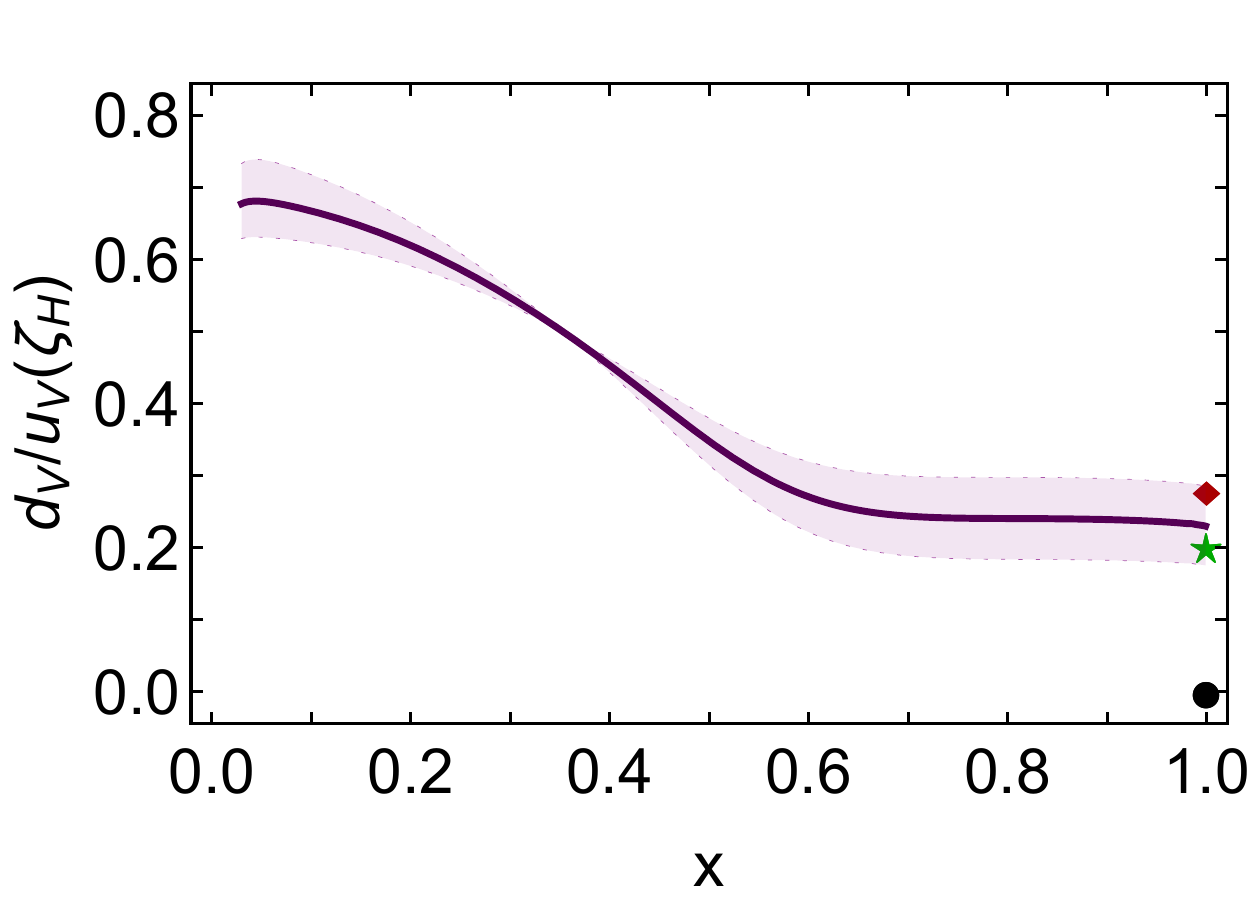}
\caption{\label{FigvalenceDfs}
\emph{Upper panel}\,--\,{\sf A}.
Proton valence-quark DFs, evaluated using ${\mathpzc r}_{10}=0.47$: ${\mathpzc u}^p(x;\zeta_{\cal H})$ -- solid red curve; ${\mathpzc d}^p(x;\zeta_{\cal H})$ -- dot-dashed blue curve; and $\tfrac{1}{2}{\mathpzc u}^p(x;\zeta_{\cal H})$ -- dashed green curve.
\emph{Lower panel}\,--\,{\sf B}.
${\mathpzc d}^p(x;\zeta_{\cal H})/{\mathpzc u}^p(x;\zeta_{\cal H})$ obtained using ${\mathpzc r}_{10}=0.47(6)$, which yields the large-$x$ value in Eq.\,\eqref{donux1}.
Other predictions:
green star -- helicity conservation in the QCD parton model \cite{Farrar:1975yb, Brodsky:1979gy, Brodsky:1994kg};
red diamond -- continuum Schwinger function methods \cite{Roberts:2013mja}.
Excluding pseudovector diquarks from the proton produces a large-$x$ value for this ratio that lies in the neighbourhood of the filled circle \cite{Close:1988br, Xu:2015kta}.
Momentum fractions associated with the DFs drawn:
$\langle x \rangle_{{\mathpzc u}_p}^{\zeta_{\cal H}} = 0.687(5)$,
$\langle x \rangle_{{\mathpzc d}_p}^{\zeta_{\cal H}} = 0.313(5)$.
}
\end{figure}

Evaluation of the hadron-scale proton valence-quark DFs is now a straightforward numerical exercise.  The results are drawn in Fig.\,\ref{FigvalenceDfs}A.
%
%
The effect of diquark correlations in the proton Faddeev amplitude is highlighted by Fig.\,\ref{FigvalenceDfs}B, which depicts the ratio of these valence-quark DFs as evaluated with ${\mathpzc r}_{10}=0.47(6)$:
\begin{equation}
\label{donux1}
\lim_{x\to 1}{\mathpzc d}_V^p(x;\zeta_{\cal H})/{\mathpzc u}_V^p(x;\zeta_{\cal H})
= 0.230(55)\,.
\end{equation}
The ${\mathpzc r}_{10}$ range was chosen such that the band of values matches that extracted \cite{Cui:2021gzg} from contemporary experiment \cite[MARATHON]{Abrams:2021xum} and data analyses \cite{Segarra:2019gbp}.
The band in Eq.\,\eqref{donux1} includes the value obtained using continuum Schwinger function methods \cite{Roberts:2013mja, Segovia:2014aza, Xu:2015kta} and also helicity conservation in the context of the QCD parton model \cite{Farrar:1975yb, Brodsky:1979gy, Brodsky:1994kg}.
It excludes a proton constituted solely as quark+scalar-diquark \cite{Close:1988br}.

\medskip

\noindent\textbf{6.\,DF evolution}.
Whilst the $x=1$ result in Eq.\,\eqref{donux1} is a fixed point under QCD evolution \cite{Holt:2010vj, Cui:2021sfu}, at any other value of $x\in (0,1)$, the DFs evolve as $\zeta$ is increased above $\zeta_{\cal H}$.  We implement evolution by using the all-orders scheme introduced elsewhere \cite{Ding:2019lwe, Cui:2020tdf, Chang:2021utv} and elucidated in Refs.\,\cite{Raya:2021zrz, Cui:2021mom, Cui:2022bxn}.
Our practical implementation is based on the process-independent effective charge, $\hat{\alpha}(k^2)$, described in Refs.\,\cite{Binosi:2016nme, Cui:2019dwv, Roberts:2021nhw}.  This charge agrees to better than 0.1\% with QCD's one-loop coupling on $k^2 \gtrsim 4 m_p^2$; but as $k^2$ continues to run toward zero, $\hat{\alpha}(k^2)$ exhibits a qualitative change.  Its behaviour on
\begin{equation}
\label{EqmG}
k^2 < \zeta_{\cal H}^2 := (0.331(2)\,{\rm GeV})^2\,,
\end{equation}
reveals that gluon modes with $k^2 \lesssim \zeta_{\cal H}^2$ are screened from interactions and QCD has entered a practically conformal domain.
The value of $\zeta_{\cal H}^2$ is fixed by QCD's renormalisation group invariant gluon mass \cite{Binosi:2016nme, Cui:2019dwv, Roberts:2021nhw}.
Given the properties of $\hat{\alpha}(k^2)$, the all-orders approach is equivalent to standard DGLAP evolution on any domain for which perturbative QCD is applicable.

Notwithstanding these qualities of $\hat{\alpha}(k^2)$, it is worth recalling that the pointwise form is largely immaterial.  As highlighted elsewhere \cite{Cui:2021mom, Cui:2022bxn}, any effective charge which supports a DF evolution scheme that is all-orders exact will produce evolved distributions with the same character.

Owing to Eq.\,\eqref{momentumsumrule}, the proton's sea and glue DFs are identically zero at $\zeta=\zeta_{\cal H}$.  
They are generated via evolution on $\zeta>\zeta_{\cal H}$ through gluon emission from the proton's dressed valence constituents in the process of shedding their clothing.  The character of these sea and glue distributions is therefore determined by the nonperturbative information contained in ${\mathpzc d}_V^p(x;\zeta_{\cal H})$, ${\mathpzc u}_V^p(x;\zeta_{\cal H})$.

Using the usual evolution kernel, gluon splitting produces quark+antiquark pairs of all flavours with equal likelihood.  However, as the proton contains two valence $u$-quarks and one valence $d$-quark, it was long ago argued \cite{Field:1976ve} that the Pauli exclusion principle should instead force gluon splitting to favour $d+\bar d$ production over $u+\bar u$.  No prediction for the resulting modification is available; so, we implement such Pauli blocking via a simple modification of the gluon splitting function:
\begin{equation}
\label{gluonsplit}
P_{f\bar f \leftarrow g}(x;\zeta) \to P_{f\bar f \leftarrow g}(x) +
 \sqrt{3}  (1 - 2 x) \frac{ {\mathpzc g}_{f\bar f} }{1+(\zeta/\zeta_H-1)^2}\,,
\end{equation}
where $P_{f\bar f \leftarrow g}(x) $ is the standard one-loop gluon splitting function, ${\mathpzc g}_{s\bar s}=0={\mathpzc g}_{c\bar c}$, and ${\mathpzc g}_{d\bar d}= -{\mathpzc g}_{u\bar u}=:{\mathpzc g}$ is a strength parameter.
This Pauli blocking term preserves baryon number; shifts momentum into the $d+\bar d$ DF from the $u+\bar u$ DF whilst maintaining the total sea momentum fraction; and vanishes with increasing $\zeta$, reflecting the diminishing influence of valence-quarks as the proton's glue and sea content increases.
With ${\mathpzc g}=1$, a kindred modification in connection with the $K$-meson acts to suppress the $s\bar s$ sea momentum fraction with respect to that of the $u\bar u$ sea by $\lesssim 2$\% at $\zeta = 5.2\,$GeV \cite{Cui:2020tdf}.

\medskip

\begin{figure}[t]
\vspace*{0.5ex}

\leftline{\hspace*{0.5em}{\large{\textsf{A}}}}
\vspace*{-3ex}
\includegraphics[width=0.42\textwidth]{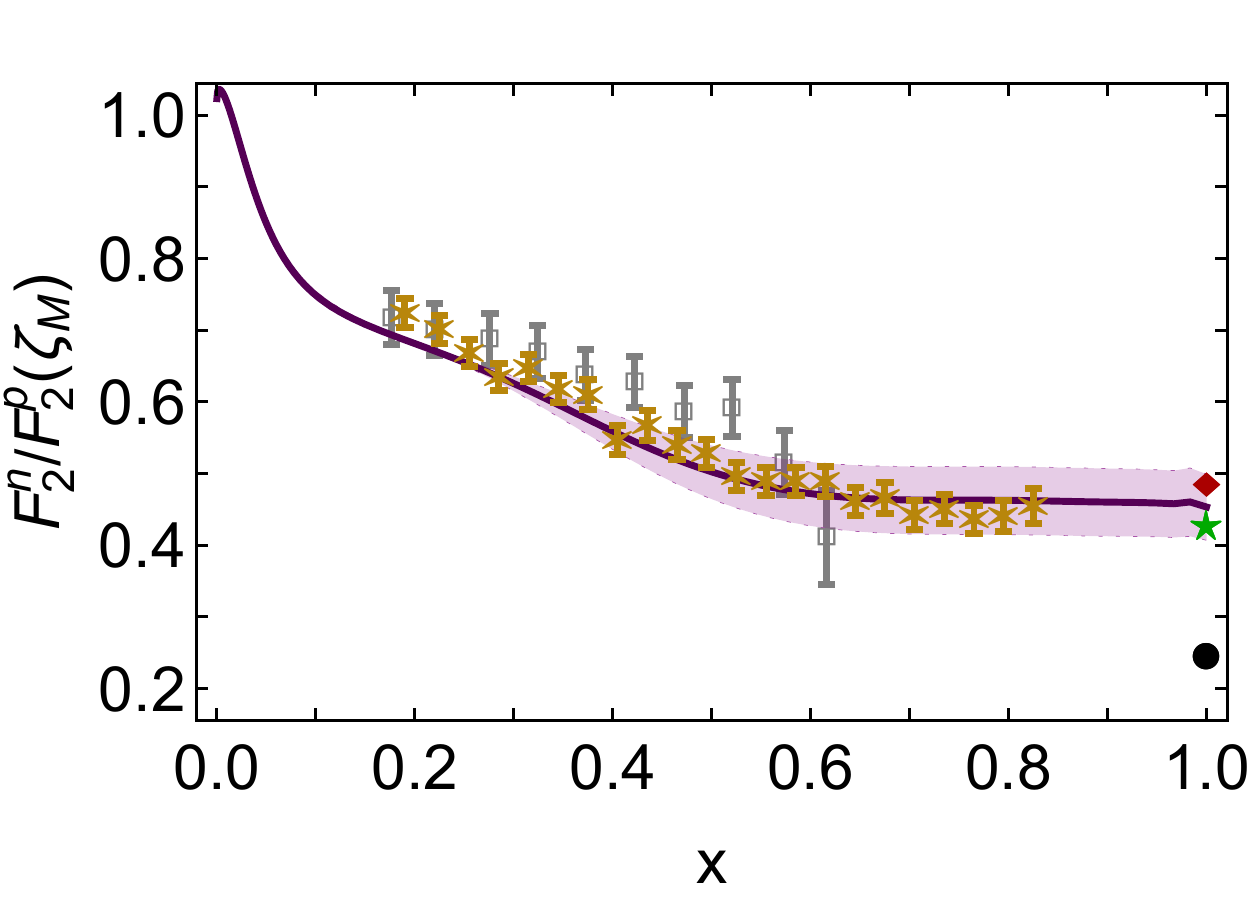}
\vspace*{-1ex}

\leftline{\hspace*{0.5em}{\large{\textsf{B}}}}
\vspace*{-3ex}
\includegraphics[width=0.42\textwidth]{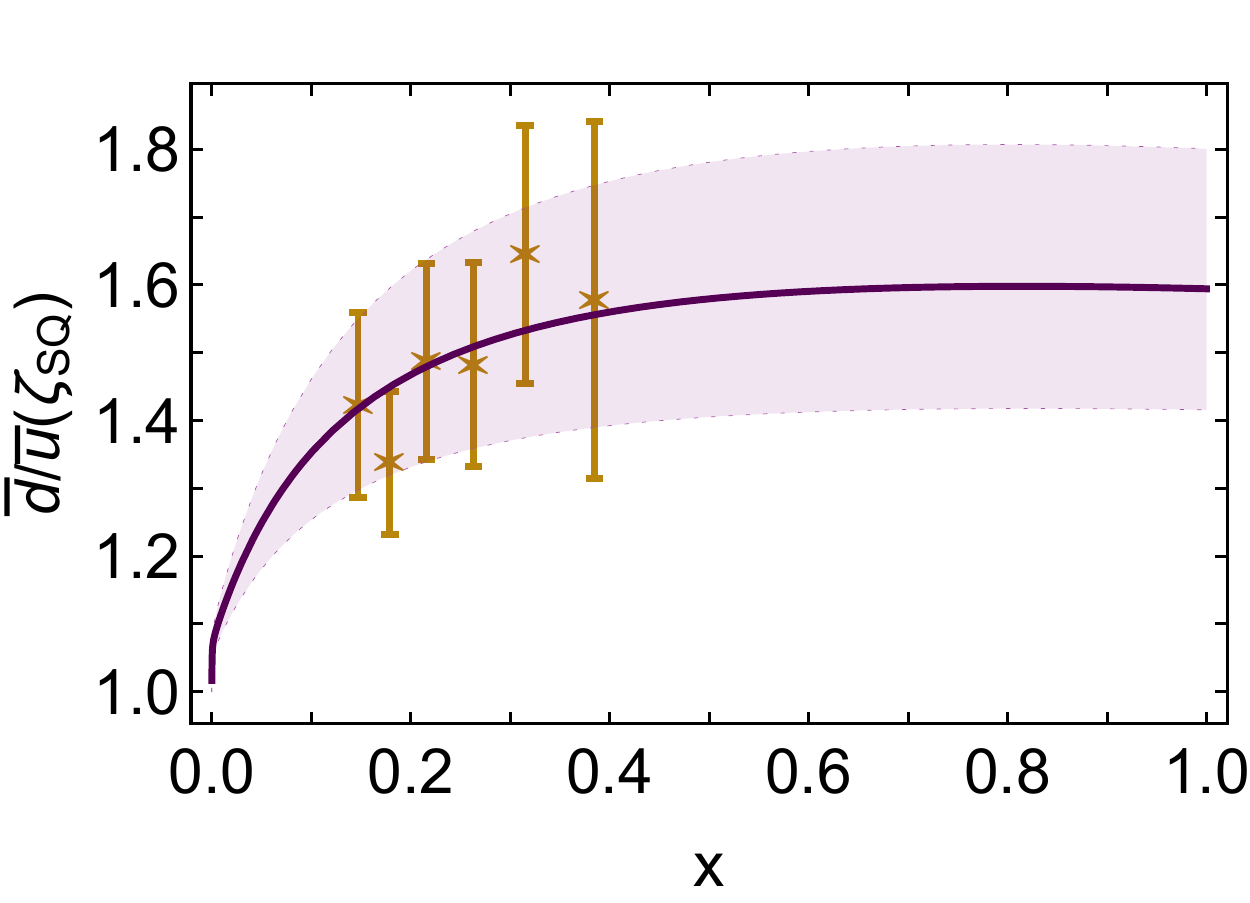}
\caption{\label{FigEvolved}
\emph{Upper panel}\,--\,{\sf A}.
Neutron-to-proton structure function ratio.
Data: open grey squares \cite[BoNuS]{CLAS:2014jvt}; and gold asterisks \cite[MARATHON]{Abrams:2021xum}.
Solid purple curve: result obtained from the valence-quark DFs in Fig.\,\ref{FigvalenceDfs}A after evolution to $\zeta^2=\zeta_{\rm M}^2 = 7.2\,$GeV$^2$.
Associated band expresses response to changes in nucleon pseudovector diquark content.
Other predictions:
red diamond -- continuum Schwinger function methods \cite{Roberts:2013mja};
green star -- helicity conservation in the QCD parton model \cite{Farrar:1975yb, Brodsky:1979gy, Brodsky:1994kg};
and retaining only scalar diquarks in the proton wave function, which produces a large-$x$ value for this ratio that lies in the neighbourhood of the filled circle \cite{Close:1988br, Xu:2015kta}.
\emph{Lower panel}\,--\,{\sf B}.
Ratio of light antiquark DFs.
Data from Ref.\,\cite[E906]{SeaQuest:2021zxb}.
Solid purple curve: result obtained from the valence-quark DFs in Fig.\,\ref{FigvalenceDfs}A after evolution to $\zeta^2=\zeta_{\rm SQ}^2 = 30\,$GeV$^2$.
Associated band expresses response to variations ${\mathpzc g}=0.4 \pm 0.1$ in Eq.\,\eqref{gluonsplit}.
}
\end{figure}

\noindent\textbf{7.\,Neutron-proton structure function ratio}.
On any domain for which an interpretation in terms of proton DFs is valid, the neutron-proton structure function ratio can be written:
\begin{align}
\label{F2nF2p}
\frac{F_2^n(x)}{F_2^p(x)} =
\frac{
{\mathpzc U}(x;\zeta) + 4 {\mathpzc D}(x;\zeta) + \Sigma(x;\zeta)}
{4{\mathpzc U}(x;\zeta) + {\mathpzc D}(x;\zeta) + \Sigma(x;\zeta)}\,,
\end{align}
where, in terms of quark and antiquark DFs,
${\mathpzc U}(x;\zeta) = {\mathpzc u}(x;\zeta)+\bar {\mathpzc u}(x;\zeta)$,
${\mathpzc D}(x;\zeta) = {\mathpzc d}(x;\zeta)+\bar {\mathpzc d}(x;\zeta)$,
$\Sigma(x;\zeta) = {\mathpzc s}(x;\zeta)+\bar {\mathpzc s}(x;\zeta)
  +{\mathpzc c}(x;\zeta)+\bar {\mathpzc c}(x;\zeta)$.
A measurement of this ratio requires an effective neutron target.  Following Refs.\,\cite{Bodek:1973dy, Poucher:1973rg}, many experiments have used the deuteron.  However, despite being a weakly bound system, numerous nuclear physics effects introduce a large theory uncertainty in the extracted ratio on $x\gtrsim 0.7$ \cite{Whitlow:1991uw, Arrington:2021vuu}.
It has been argued \cite{Afnan:2000uh, Pace:2001cm} that one may instead obtain a sound result for $F_2^n(x)/F_2^p(x)$ by forming the ratio of deep inelastic scattering measurements on $^3$He and $^3$H.  Data from such an experiment are now available \cite{Abrams:2021xum}.  They are reproduced in Fig.\,\ref{FigEvolved}A and may be referred to a scale $\zeta^2=\zeta_M^2 = 7.2\,$GeV$^2$.
(The results in this section are independent of ${\mathpzc g}$ in Eq.\,\eqref{gluonsplit}.)

\begin{table}[t]
\caption{
\label{tabmoments}
Light-front momentum fractions, $\langle x \rangle$, for each parton class in the proton and pion.
Concerning $\pi$, ``${\mathpzc d}$'' should be interpreted as ``$\bar {\mathpzc d}$''.
($\zeta_{\rm M}^2=7.2\,$GeV$^2$ and $\zeta_{\rm SQ}^2=30\,$GeV$^2$.)
}
\begin{tabular*}
{\hsize}
{
l@{\extracolsep{0ptplus1fil}}
|c@{\extracolsep{0ptplus1fil}}
c@{\extracolsep{0ptplus1fil}}
c@{\extracolsep{0ptplus1fil}}
c@{\extracolsep{0ptplus1fil}}
c@{\extracolsep{0ptplus1fil}}
c@{\extracolsep{0ptplus1fil}}}\hline
system$\ $ & ${\mathpzc u}$ & ${\mathpzc d}$ & total valence & sea & glue \\\hline
$p_{\zeta_{\rm M}}\ $  & $0.31\ $ & $0.14\ $& $0.45\ $ & $0.12\ $ & $0.43\ $\\
$p_{\zeta_{\rm SQ}}\ $ & $0.28\ $ & $0.13\ $& $0.41\ $ & $0.14\ $ & $0.45\ $\\
$\pi_{\zeta_{\rm SQ}}\ $ & $0.21\ $ & $0.21\ $& $0.41\ $ & $0.14\ $ & $0.45\ $\\\hline
\end{tabular*}
\end{table}

Beginning with the valence-quark DFs in Fig.\,\ref{FigvalenceDfs}A and evolving as described in Sect.\,6, one obtains the solid curve drawn in Fig.\,\ref{FigEvolved}A.  The related band expresses the variation ${\mathpzc r}_{10}=0.47(6)$ in relative pseudovector:scalar diquark strength; and
\begin{equation}
\lim_{x\to 1} F_2^n(x)/F_2^p(x)= 0.453(46)\,.
\end{equation}
Consistent with the analysis in Ref.\,\cite{Cui:2021gzg}, the calculated result agrees well with the data, which is thus seen to strongly disfavour a proton wave function that is based solely on scalar diquark correlations, \emph{i.e}., excludes pseudovector diquarks.

The behaviour of the curve in Fig.\,\ref{FigEvolved}A is determined by the valence-quark DFs on $x\gtrsim 0.2$.  However, sea-quark contributions become increasingly important on the complementary domain, especially as $x\to 0$; and this explains the associated saturation: $\lim_{x\to 0} F_2^n(x)/F_2^p(x) \simeq 1$.
The momentum fractions are listed in Table~\ref{tabmoments}
%
%
and agree with an array of estimates based on fits to a wide variety of high-precision data \cite[Table\,VI]{Hou:2019efy}.

\medskip

\noindent\textbf{8.\,Asymmetry of light-quark sea}.
Expressed via neutron and proton structure functions, measured using charged-lepton (${\mathpzc l}$) beams, the Gottfried sum rule relates to the following integral \cite{Gottfried:1967kk, Brock:1993sz}:
\begin{equation}
{\mathpzc S}_G = \int_0^1 \frac{dx}{x}\,[F_2^{{\mathpzc l} p}(x) - F_2^{{\mathpzc l} n}(x)]\,.
\end{equation}
Assuming isospin symmetry, then on any kinematic domain for which one may express the structure functions in terms of DFs:
\begin{equation}
\label{Gottfried}
{\mathpzc S}_G  = \frac{1}{3} - \frac{2}{3}{\mathpzc I}_{\bar d - \bar u}^{0,1}\,,
\quad
{\mathpzc I}_{\bar d - \bar u}^{0,1}  = \int_0^1 dx\,[\bar {\mathpzc d}(x) - \bar {\mathpzc u}(x)]\,,
\end{equation}
where Eqs.\,\eqref{baryonnumber} have been used.
Experiments suggest \cite{NewMuon:1991hlj, NewMuon:1993oys, NA51:1994xrz, NuSea:2001idv, SeaQuest:2021zxb} that ${\mathpzc I}_{\bar d - \bar u} >0$, \emph{viz}.\ the proton's light-quark sea exhibits an antimatter asymmetry.

On domains for which the steps to Eq.\,\eqref{Gottfried} are valid, momentum conservation entails $\int_0^1 dx\,x \{\bar{\mathpzc d},\bar{\mathpzc u}\}<\infty$.  Typical data fits preserve this result but indicate $\int_0^1 dx\,\{\bar{\mathpzc d},\bar{\mathpzc u}\}=\infty$ \cite{Hou:2019efy}, where, given that all DFs vanish smoothly as $x\to 1$, the divergence is associated with sea DF behaviour on $x\simeq 0$.  These characteristics are also seen in continuum predictions of sea (and glue) parton DFs \cite{Ding:2019lwe, Cui:2020tdf, Chang:2021utv}. Consequently, ${\mathpzc I}_{\bar d - \bar u} <\infty$ is only possible if the leading $x\simeq 0$ divergence in each antiquark DF cancels to leave a finite difference, or, at worst, the subtraction leaves a singularity that is integrable on any domain containing $x=0$.  This is why data fits using unconstrained functions do not yield a defined value for ${\mathpzc I}_{\bar d - \bar u}^{0,1}$ \cite{NNPDF:2017mvq}.
%

The $x$-dependence of the antimatter asymmetry indicated by ${\mathpzc I}_{\bar d - \bar u} >0$ has been exposed in data that can be related to the ratio $\bar d(x;\zeta)/\bar u(x;\zeta)$ \cite{NuSea:2001idv, SeaQuest:2021zxb}.  The most recent set is displayed in Fig.\,\ref{FigEvolved}B.  It can be referred to a scale $\zeta^2=\zeta_{\rm SQ}^2=30\,$GeV$^2$.

Working with the valence-quark DFs in Fig.\,\ref{FigvalenceDfs}A and evolving as described in Sect.\,6, one obtains the solid curve drawn in Fig.\,\ref{FigEvolved}B: the central curve and band correspond to ${\mathpzc g}=0.4(1)$ in Eq.\,\eqref{gluonsplit}.  Using these results, one obtains
\begin{equation}
{\mathpzc I}_{\bar d - \bar u}^{0.004,0.8} = 0.091(23)
\end{equation}
for the Gottfried sum rule discrepancy on the domain covered by Refs.\,\cite{NewMuon:1991hlj, NewMuon:1993oys}.  Our result may be compared with the value inferred from recent fits to a wide variety of high-precision data \cite[CT18]{Hou:2019efy}: 0.110(80).
Extending the integral upper bound has very little impact; but taking the lower bound to zero, our calculated DFs yield ${\mathpzc I}_{\bar d - \bar u}^{0,1} = 0.52(12)$, highlighting the importance of the domain $x\lesssim 0.001$.

The DFs used to calculate the results displayed in Fig.\,\ref{FigEvolved}B produce the momentum fractions listed in Table~\ref{tabmoments}\,--\,Row~2.
%
%
%
%
It is worth adding that
\begin{equation}
\langle x \rangle_{\bar{\mathpzc d}_p}^{\zeta_{\rm SQ}}
-\langle x \rangle_{\bar{\mathpzc u}_p}^{\zeta_{\rm SQ}} = 0.0029(7)\,.
\end{equation}
This is a relative shift of just $\approx 4$\% between the momentum fractions carried by the light antiquarks in the sea, an outcome which emphasises that the seemingly large deviation of $\bar d(x;\zeta_{\rm SQ})/\bar u(x;\zeta_{\rm SQ})$ from unity, evident in Fig.\,\ref{FigEvolved}B, can be achieved by a modest Pauli blocking term.

\medskip

\noindent\textbf{9.\,Contrasting proton and pion DFs}.
The complete array of pion DFs was computed and discussed in Refs.\,\cite{Ding:2019lwe, Cui:2020tdf, Chang:2021utv}.  These results are the basis for the comparison provided between proton and pion light-front momentum fraction decompositions in Table~\ref{tabmoments}: Row~2 \emph{cf}.\ Row~3.

At the hadron scale, $\zeta_{\cal H}$, all the proton's light-front momentum is carried by the three dressed valence quarks, Eq.\,\eqref{momentumsumrule}.  This is also true in the pion, \emph{viz}.\ the pion's two dressed valence constituents carry all the pion's momentum at $\zeta_{\cal H}$.  The results in Table~\ref{tabmoments} suggest this entails that for any $\zeta>\zeta_{\cal H}$ the light-front momentum fractions carried by identifiable parton classes are the same for the proton and pion.  On the other hand, the pointwise behaviour of the DFs is hadron-dependent.
%
%
In fact, comparing comparing Fig.\,\ref{FigvalenceDfs}A with Ref.\,\cite[Fig.\,5]{Ding:2019lwe}, such differences are seen to be great.  Two features are of particular importance: (\emph{i}) ${\mathpzc u}^\pi(x;\zeta_{\cal H})$ is far more dilated than ${\mathpzc u}_V^p(x;\zeta_{\cal H})$, ${\mathpzc d}_V^p(x;\zeta_{\cal H})$; and (\emph{ii}) the endpoint behaviour of pion valence DFs is much harder than that of the analogous proton DFs, e.g., Ref.\,\cite[Eq.\,(2)]{Ding:2019lwe} \emph{cf}.\ Eq.\,\eqref{partonmodel} above.

Such disparities between proton and pion structure likely have their origin in dissimilarities between the expressions of emergent hadron mass in the proton and pion.  Explaining them may deliver insights that can lead to an understanding of the origin of the proton mass and the character of Nambu-Goldstone bosons \cite{Roberts:2016vyn, Roberts:2021xnz}.  A detailed analysis of pointwise differences between proton and pion DFs is underway.

Owing largely to persisting controversies that surround the behaviour of pion DFs \cite{Chang:2021utv, Cui:2021mom, Cui:2022bxn}, the verity of the predictions described in this section cannot now be established.  An array of new experiments aims to lift this fog \cite{Gautheron:2010wva, JlabTDIS1, JlabTDIS2, Adams:2018pwt, Aguilar:2019teb, Chen:2020ijn, Arrington:2021biu}.

\medskip

\noindent\textbf{10.\,Summary and Perspective}.
Using an algebraic representation of the proton Faddeev amplitude, which admits the presence of nonpointlike isoscalar-scalar and isovector-pseudovector quark+quark (diquark) correlations [Sect.\,4], we calculated all proton distribution functions (DFs -- valence, sea, and glue) and presented a unified description of the neutron-proton structure function ratio [$F_2^n/F_2^p$, Sect.\,7] and the $\bar d$-$\bar u$ asymmetry in the proton [Sect.\,8].

A description of modern data \cite[MARATHON]{Abrams:2021xum} and analyses \cite{Segarra:2019gbp} relating to $F_2^n/F_2^p$ requires the inclusion of pseudovector diquarks, whose presence accounts for roughly 35\% of the proton's wave function normalisation.  A proton built solely using scalar diquark correlations is excluded by the data.
Regarding the proton sea-quark DFs, a modest Pauli blocking effect, implemented as a modification of the gluon splitting function, is sufficient to explain the observed antimatter asymmetry, delivering results in agreement with contemporary data \cite[E906]{SeaQuest:2021zxb}.

Having thus established that our \emph{Ansatz} for the Poincar\'e-covariant proton wave function is reasonable, we compared its predictions for the light-front momentum fraction decomposition over parton classes with those in the pion [Sect.\,9].  Notably, the light-front momentum fractions carried by identifiable parton classes are the same for the proton and pion, but higher moments of these DFs are different.


A natural next step in this body of analyses is a detailed comparison between the pointwise behaviour of proton and pion DFs.  That is underway \cite{Lu:2022}.
It is also desirable to repeat the analysis described herein using a refined representation of the proton's Poincar\'e-covariant wave function, e.g., including running masses for the hadron-scale dressed-constituents and additional components of the Faddeev amplitude.
A longer term goal is further development of the explicitly three-body Faddeev equation approach to proton structure so that it can be employed in the calculation of proton DFs, perhaps by exploiting, \emph{inter alia}, the developments reported in Ref.\,\cite{Eichmann:2021vnj}.

\medskip
\noindent\textbf{Acknowledgments}.
We are grateful for constructive comments from D.~Binosi, C.~Chen, Z.-F.~Cui, M.~Ding, R.\,J.~Holt, Y.-X.~Liu, J.~Papavassiliou, P.\,E.~Reimer, K.~Raya, J.~Rodr\'{\i}guez-Quintero and S.\,M.~Schmidt.
Work supported by:
National Natural Science Foundation of China (grant no.\,12135007).



\end{document}